\documentclass[11pt]{article}
\usepackage{moriond,epsfig}
\usepackage{xspace}

\def\meanPt{$\left<p_T\right>\:$}
\newcommand{\lam}{\ensuremath{\Lambda} \xspace}

\newcommand{\kz}{\ensuremath{\mathrm{K^{0}_{S}}}\xspace}
\newcommand{\XI}{\ensuremath{\Xi^{-}}\xspace}

\newcommand{\nch}{$N_{ch}$\xspace}

\newcommand{\sqs}{\ensuremath{\sqrt{s}}\xspace}
\newcommand{\kT}{\ensuremath{k_{T}}\xspace}

\newcommand{\pp}{\ensuremath{p+p}\xspace}
\newcommand{\pt}{\ensuremath{p_{T}}\xspace}

\begin{document}
\vspace*{4cm}
\title{STRANGE AND MULTI-STRANGE PARTICLE PRODUCTION IN P+P COLLISIONS AT $\sqrt{s}$=200 GeV IN STAR}

\author{Mark T.Heinz }

\address{Laboratory of High Energy Physics\\
 University of Bern\\
 3012 Bern, Switzerland}

\maketitle\abstracts{ We present measurements of the transverse
momentum spectra for \kz, \lam, \XI and their antiparticles in p+p
collisions at $\sqs=200$ GeV. The extracted mid-rapidity yields and
\meanPt are in agreement with previous $p+\bar{p}$ experiments while
they have smaller statistical errors. We compare the measured
particle spectra and \meanPt values to predictions from the PYTHIA
leading order pQCD model (v6.221) and see significant disagreements
with the default settings. Finally we compare the spectra to the
latest calculations from NLO pQCD and see a good agreement for the
\kz, but large discrepancies \lam.}

\section{Introduction}
We report results from the 2002 \pp running with the STAR experiment
at RHIC. We present the high statistics measurement of \kz,
$\Lambda$ and $\Xi$ particles and obtain the yield and \meanPt for
each species. We compare our measurement to the PYTHIA model
incorporating leading-order (LO) pQCD processes and note large
discrepancies between our data and the model in it's default
setting. Attempts to tune the LO K-Factor and the intrinsic parton
momentum distribution \kT produce better agreements. We also compare
to next-to-leading order (NLO) calculations which exhibit a better
agreement with our \kz data, but still fail to describe the
$\Lambda$. Furthermore, we study the dependency of \meanPt as a
function of charged particle event multiplicity ($N_{ch}$) for
different particle species. It has been shown that the high
transverse momentum final state particle is mostly governed by hard
partonic processes and subsequent string fragmentation \cite{Sjo87}.
We observe a strong dependence of this part of the spectra with
respect to event multiplicity.

\section{Analysis Description}
The present data were reconstructed using the STAR experiment which
is described in more detail elsewhere \cite{STAR2}. The main
detector used in this analysis is the Time Projection Chamber (TPC)
covering the full acceptance in azimuth together with a large
pseudo-rapidity coverage ($\mid \eta \mid < 1.5$). A total of 14
million non-singly diffractive (NSD) events were triggered with the
STAR beam-beam counters (BBC) requiring two coincident charged
tracks at forward rapidity. Due to the particulary low track
multiplicity environment in p+p collisions only 76\% of primary
vertices are found correctly; from the remainder, 14\% are lost and
10\% are incorrectly reconstructed as demonstrated by a MC-study. Of
all triggered events, 11 million events passed the selection
criteria requiring a valid primary vertex within 100cm along the
beam-line from the center of the TPC detector. The strange particles
were identified from their weak decay to charged daughter particles.
The following decay channels and the corresponding anti-particles
were analyzed: \kz $\rightarrow \pi^{+} + \pi^{-}$ (b.r. 68.6\%),
$\Lambda \rightarrow p + \pi^{-}$(b.r. 63.9\%) ,$\Xi^{-} \rightarrow
\Lambda + \pi^{-}$(b.r. 99.9\%). Particle identification of
daughters was achieved by requiring the dE/dx to fall within the
3$\sigma$-bands of theoretical Bethe-Bloch parameterizations.
Further background in the invariant mass was removed by applying
topological cuts to the decay geometry. Corrections for acceptance
and particle reconstruction efficiency were obtained by a
Monte-Carlo based method of embedding simulated particle decays into
real events and comparing the number of simulated and reconstructed
particles in each $p_{T}$-bin.

\section{Results for mid-rapidity yield and \meanPt}

\begin{table}[ht]
\begin{center}
\begin{tabular}{|c|c|c|c|c|c|}
\hline Particle  & dN/dy  $\mid$y$\mid <$ 0.5  & \meanPt [GeV/c] \\
\hline \kz       & 0.128 $\pm$0.002(stat) $\pm$0.010(syst)   & 0.60 $\pm$0.01(stat)$\pm$0.02(syst) \\
\hline $\Lambda + \bar{\Lambda}$ & 0.070 $\pm$0.002(stat) $\pm$0.006(syst)   & 0.76 $\pm$0.02(stat)$\pm$0.04(syst) \\
\hline $\Xi + \bar{\Xi}$         & 0.0036 $\pm$0.0002(stat) $\pm$0.0010(syst)& 0.96 $\pm$0.05(stat)$\pm$0.09(syst) \\
\hline
\end{tabular}
\caption {Results of mid-rapidity yields and $\langle \mathrm{p_{T}}
\rangle$ for \kz, $\Lambda$ (feed-down corrected) and $\Xi$ and
measured in \pp collisions at \sqs=200 GeV} \label{tab:Results}
\end{center}
\end{table}

In table \ref{tab:Results} results for yield and \meanPt of
corrected inclusive spectra are shown for \kz, $\Lambda$, $\Xi$ and
their respective antiparticles. The particle acceptance at
mid-rapidity ($\mid$y$\mid$ $\leq$ 0.5) in the TPC starts at a
transverse momentum of 0.2 GeV/c for \kz, 0.3 GeV/c for the
$\Lambda$ and 0.5 GeV/c for the $\Xi$. In order to extract the
$\langle \mathrm{p_{T}} \rangle$ and yield at mid-rapidity, a
parametrization to the spectra has to be applied to extrapolate the
measurement to cover the full $\mathrm{p_{T}}$-range. In contrast to
previous $p+\bar{p}$ experiments, which used either a single
exponential function in transverse mass or power-law functions, we
found that a combination of these functions is more effective in
fitting the singly-strange particles. Composite fits, using an
exponential function in $m_{T}$ at low $\mathrm{p_{T}}$ and
power-law functions at high $\mathrm{p_{T}}$ yielded the lowest
$\chi^{2}$ and were used to extract the values for yield and
$\langle \mathrm{p_{T}} \rangle$ of \kz and $\Lambda$. For the $\Xi$
spectrum the limited coverage at low $\mathrm{p_{T}}$ makes it
insensitive to the different functions and thus only an exponential
function in $m_{T}$ was used. The values as shown in table
\ref{tab:Results} are in agreement with the measurements by UA5 when
scaled using a rapidity distribution obtained from simulation
\cite{MH:SQM04}.

The systematical errors are mainly due to the different fit
parameterizations as well as the normalization uncertainty due to
`pile-up'. The $\bar{\Lambda}$/$\Lambda$ ratio is 0.92$\pm$0.09 and
$\bar{\Xi}$/$\Xi$ ratio is 0.90$\pm$0.09 and both are independent of
$\mathrm{p_{T}}$. The data sample was split into event classes with
increasing mean charged particle multiplicity per unit $\eta$. For
\kz and $\Lambda$ six event classes with sufficient statistics were
possible. The goal is to study the large momentum transfer region of
the parton-parton collisions by measuring the spectra in high
multiplicity events, where this type of interaction is expected to
be more probable.

\section{Comparison to PYTHIA 6.221 (LO pQCD)}
\begin{figure}[h]
\centering \epsfig{figure=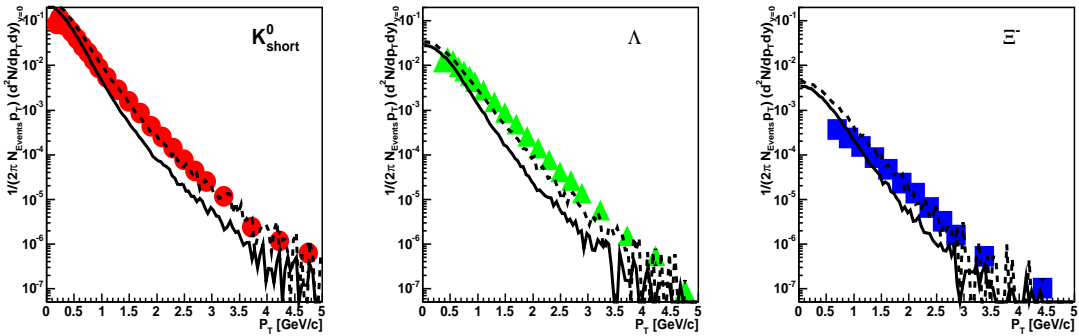,
height=5cm, width=15cm} \caption{\kz (left), \lam (center) and \XI
(right) compared to PYTHIA 6.221. Full line is default settings,
dashed line is tuned with LO K-Factor =3} \label{fig:Spectra}
\end{figure}

\begin{figure}[ht]
\centering \epsfig{figure=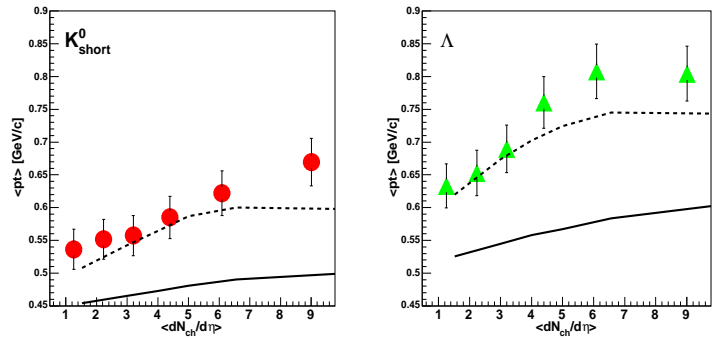, height=5cm,
width=10cm} \caption{\meanPt vs Nch for \kz (left) and \lam (right)
compared to PYTHIA 6.221. Full line is default settings, dashed line
is tuned with intrinsic $k_{T}=4$ GeV} \label{fig:MeanPt}
\end{figure}

In figure \ref{fig:Spectra} the \pt-spectra are compared to
predictions from PYTHIA LO model. The default settings of version
6.221 clearly underestimate the data. However by introducing a LO
K-Factor in agreement with studies by Eskola et al. the simulation
describes the data much better \cite{Eskola:03}. Figure
\ref{fig:MeanPt} presents the \meanPt vs $\langle dN_{ch}/d\eta
\rangle$ for \kz and \lam. A rise in \meanPt with increasing
$N_{ch}$ is observed and the increase is stronger for the $\Lambda$
than for the \kz. Several authors have attributed this phenomenon to
the increased number of large momentum transfer parton-parton
collisions that produce mini-jets in the high multiplicity events
\cite{Gyu92}. In order to reproduce the strength of the correlation
between \meanPt and \nch the intrinsic \kT value was tuned to 4 GeV,
although considered very high, from a default value of 1 GeV.

\section{Comparison to NLO pQCD calculations}

\begin{figure}[ht]
\centering \epsfig{figure=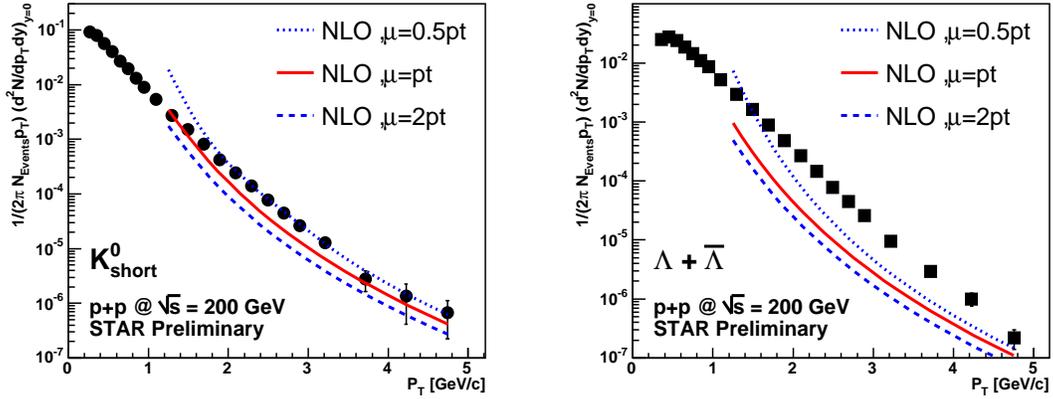, height=6cm, width=15cm}
\caption{\kz (left) and \lam(right) \pt-spectra compared to NLO
calculations using fragmentation functions by KKP for \kz and Jager
et al. for \lam.Errors shown are statistical only.} \label{fig:NLO}
\end{figure}

It has been shown previously that the charged particle and neutral
pion spectra at RHIC are well described by NLO pQCD calculations
\cite{MVL2005,David:04}. Therefore it is interesting to test the NLO
calculations for the strange particles as shown in figure
\ref{fig:NLO}. These calculations use KKP fragmentation functions
for \kz \cite{KKP} and fragmentation functions by Vogelsang et al.
for $\Lambda$ \cite{WV}. The disagreement between the calculations
and the data are small for \kz but considerably larger for the
$\Lambda$. This may be due to the higher mass of the $\Lambda$,
where the massless quark formalism breaks down and the
(m/$\mathrm{p_{T}}$)-scale approximations become non-negligible.
Bourrely and Soffer have calculated alternative fragmentation
functions for octet baryons which need to be tested \cite{Bourrely}.

\section{Summary}
The STAR experiment has made the first high statistics measurement
of mid-rapidity \kz, $\Lambda$ and $\bar{\Lambda}$, $\Xi$ and
$\bar{\Xi}$ in $p+p$ collisions at $\sqrt{s}$ = 200 GeV. The results
agree with those made by the UA5 collaboration for $p+\bar{p}$
collisions at the same energy. The ratio of
$\bar{\Lambda}$/$\Lambda$ and $\bar{\Xi}$/$\Xi$ suggests a small net
baryon number at mid-rapidity. Furthermore, we show that PYTHIA
6.221 needs to be tuned significantly to describe the strange
particle spectra in STAR. Also, NLO calculations agree reasonably
well with \kz data above 1.5 GeV/c but fail to reproduce the shape
of the $\Lambda$ spectra. Finally, we have undertaken studies to
understand the change in $\langle \mathrm{p_{T}} \rangle$ of
different strange particle species with increasing event
multiplicity in an attempt to understand the flavor dependance of
fragmenting mini-jets in high multiplicity event samples.

\end{document}